FRONT MATTER

**Title**
High throughput spectroscopy of pL droplets

**Authors**
Marc Sulliger,[1] Jaime Ortega Arroyo,[1]* Romain Quidant[1]*

**Affiliations**
[1] Nanophotonic Systems Laboratory, Department of Mechanical and Process Engineering, ETH Zurich, 8092 Zurich, Switzerland.

*Corresponding authors: jarroyo@ethz.ch, rquidant@ethz.ch

**Abstract**
Droplet microfluidics offers a versatile platform for analyzing liquid samples. Despite its potential, there is a lack of techniques that allow to reliably probe individual circulating droplets. The prospect of combining droplet microfluidics with sensitive, broadband spectroscopic techniques would therefore unlock new capabilities for various disciplines, including biomedicine and biochemistry. Here we present an integrated optofluidic platform that seamlessly combines droplet microfluidics and advanced hyperspectral imaging. This enables high-resolution, label-free analysis of single picolitre-sized droplets, providing valuable insights into their content. As a proof-of-principle, we demonstrate the ability of our platform to study rapid dynamic changes in a heterogeneous population of plasmonic nanoparticles with millisecond-time resolution. Furthermore, we demonstrate the effectiveness of the platform in biosensing applied to short DNA strands, achieving a detection sensitivity in the range of 100 pM. Finally, we show that the platform provides the flexibility to monitor samples over extended periods of time (hours) in a multiplexed manner.

**Teaser**
A platform combining droplet microfluidics, hyperspectral imaging, and measurement automation for versatile liquid sample interrogation.

# MAIN TEXT

## Introduction

State-of-the-art microfluidic devices are nowadays routinely applied to a broad range of samples – be it in the context of medicine[1–8], chemical synthesis[9–11], biochemistry[12,13], or material science[14–16]. One of the prominent examples of this is droplet microfluidics. The benefits of droplet microfluidics combined with optical readout have been shown extensively in many research articles and review papers[17–20]. In terms of the experimental procedure, droplets themselves define individual but identical reaction containers making each droplet an independent experiment. Droplets allow for high throughput measurements (kHz production frequencies are routinely achieved) and only require a small (but still easily adjustable) volume. In sensing applications, a droplet acts as a 3D sensing volume. When in motion, it generates internal flow fields that facilitate continuous sample circulation and mixing. Microfluidic droplet chips offer a diverse toolbox of design elements enabling complex on-chip functionalities. Additionally, the "closed" nature of droplets prevents contamination of the chip's region of interest (ROI), facilitating chip reusability and reducing production cost and effort.

Typical methods to analyze droplets can be categorized into two main approaches. The first approach forms the basis for digital assays. It relies on assigning a binary value to each droplet, e.g. whether a certain species is present or not, and subsequently accumulating large numbers of droplets to determine the underlying concentration of a target analyte with ultra-high sensitivity[21]. This approach often requires off-chip incubation as well as a dedicated read-out chip. More importantly digital approaches generally do not provide time-resolved information and depend on the stochastic loading of each droplet such that on average there is less than one target molecule per droplet. The second approach consists of assigning analog values to each droplet by either spectroscopic (e.g. fluorescence, photothermal or absorption/extinction spectroscopy) or electrochemical (e.g. impedance measurements) techniques. Despite enabling time-resolved or high temporal resolution studies, applications based on analog signal processing predominantly compare a single measurement value (i.e. single wavelength absorption/emission) to a threshold; therefore, binarizing the information about each droplet's content. To increase the information content, broadband spectral droplet read-out has been proposed for different spectroscopic techniques[22–27]. Whilst some of these approaches achieve considerable sensitivity, they are all limited either by the data throughput (number of spectra/s, minimum sample volume, temporal resolution), the complexity in the instrumentation (e.g. cavity enhancement), or the requirements for the chip fabrication (e.g. chip integrated optical fibers or optical path length enhancing geometries).

Hyperspectral imaging (HSI)[28], initially developed for earth remote sensing[29], is a promising approach to address the throughput limitations of current broadband spectral readout schemes, as has been recently demonstrated for the analysis of µL-volumes [30]. This is because HSI simultaneously records spatial and spectral data over a large field of view. When applied to droplet microfluidics, a single recorded image contains the information of several droplets. In combination with µs integration times this approach not only ensures high throughput and direct retrieval of the droplet length, but also enables position dependent time encoding. Furthermore, depending on the chip geometry and the droplet production frequency, access to fast reaction kinetics is possible[31]. Whilst the number of spectra per image is simply limited by the size of the camera sensor, the entire readout remains rather simple and does not require specialized cameras.

In this work we introduce an optofluidic platform that integrates state-of-the-art droplet microfluidics with HSI-based detection to interrogate pL-volume droplets with high spectral and temporal resolution. The platform achieves high throughput screening by combining the high temporal resolution of the optical readout together with the multiplexed sample handling from the microfluidic chip. Specifically, we designed a microfluidic chip integrated with push-up valves that achieves high droplet production rates with low reagent consumption. Together with the characterization of the parameter space for droplet production we present an optimized droplet analysis workflow to obtain spectral information down to the single droplet level. To showcase the potential for sample classification at high temporal resolution in the ms-regime, we resolve fast dynamical changes in sample composition induced by rapid valve switching.

Finally, using a model DNA sensor system comprising of nucleic acid functionalized gold nanoparticles, we demonstrate two potential applications of our platform: high throughput biosensing and real-time monitoring of complex samples.

## Results

**Working principle of the platform** - This work addresses the need for broad-band absorbance spectra with high spectral and temporal resolution at the single droplet level by developing an integrated optofluidic platform based on three key components depicted in Fig. 1: microfluidics, optical imaging, and automation. For microfluidics, the platform takes advantage of the high throughput and low sample volume consumption of droplet microfluidics. In addition, we integrate valves into the microfluidic chip to further increase control over the sample and to also deliver multiplexed sample readout. For optical imaging, the platform consists of a custom microscope with multiple channel readouts, with HSI as the main detection channel. The HSI approach simultaneously extracts spatial and spectral information within a single measurement. Compared to confocal analogs, HSI delivers higher information content per read-out, as each row of the sensor records a spectrum of the sample. Finally, for automation, the entire platform including pressure controllers, electronic valve actuation, microscope control and data acquisition are programmatically operated and synchronized. This measurement automation not only enables complex experimental protocols, but also allows for observation of very short, exactly timed sequences, and long-term monitoring (hours) of slow dynamical changes within a specific sample.

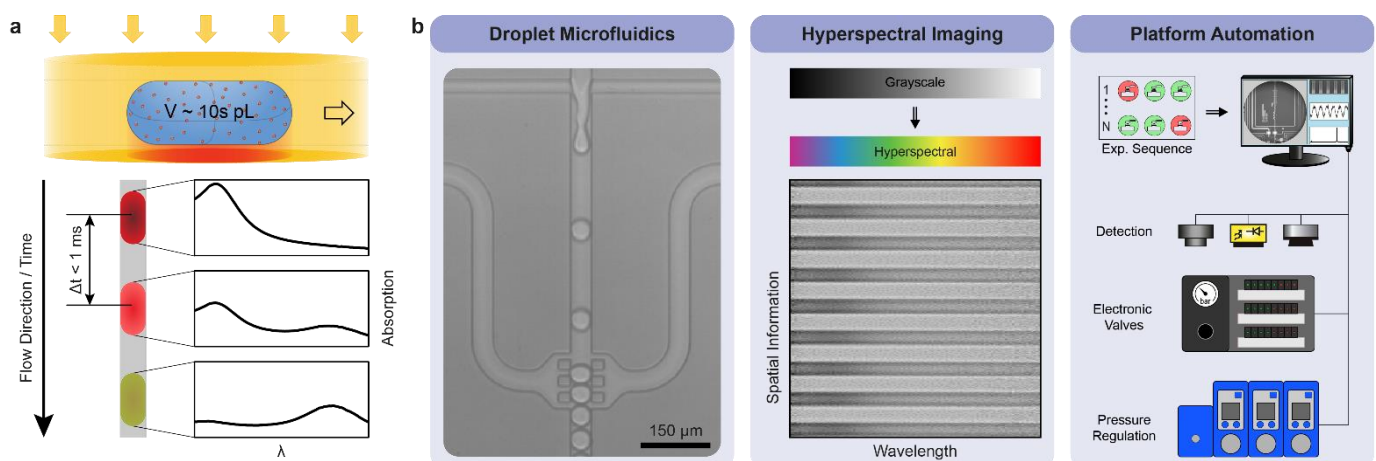

**Fig. 1. Concept and main building blocks of the experimental platform.** (a) Conceptual illustration of the platform. (b) The platform is comprised of three main building blocks: a droplet microfluidics toolbox integrated with micro valves, a custom-built microscope for label free HSI with high spectral resolution, and an automated measurement and droplet analysis workflow.

**Micro-valves integration delivers enhanced control and multiplexing to droplet microfluidics** - Our microfluidics chips (see Fig. 2a) are designed to integrate droplet production and read-out (i.e. ROI) within the same device. They are based on a two-layer architecture with microfluidic Quake valves[32] in a push up configuration[33,34]. The flow layer features a total of seven independently addressable aqueous/sample inlets as well as a waste outlet. All the sample inlets connect to a single channel before a flow focusing junction. The flow focusing junction generates the droplets by combining the oil (coming from top and bottom with a common inlet at the right-hand side of the chip) and aqueous phases. The produced droplets flow into the outlet channel, designed to have a 5-fold lower hydrodynamic resistance than each of the inlet channels. Along the outlet channel, magnified in Fig. 2b, the droplets pass through a short section composed of three meandric turns, which achieve reagent mixing by breaking the axial symmetry of the droplet internal flow patterns[35]. The outlet channel then expands by 10 µm (from 35 µm to 45 µm) before reaching the measurement ROI, which corresponds to a 1 mm long channel section. For the control layer, the chip has nine independent inlets which overlap with specific areas (Fig. 2c) of the flow channels to form the physical valves shown in Fig. 2d. Integration of micro-valves massively increases the operational control of the chip. Namely, they enable multiplexed and fully automated operation of the sample inlets, controlled sample mixing, and completely prevent reagent counterflow. There are two main reasons to choose push-up valves over traditional push-down valves. Firstly, push-up valves can close low aspect ratio channels at lower applied pressures[33]. Secondly, placing the control layer beneath the flow channel creates a uniform channel surface entirely made of polydimethylsiloxane (PDMS), which eliminates the need for hydrophobic surface treatment of the flow channel.

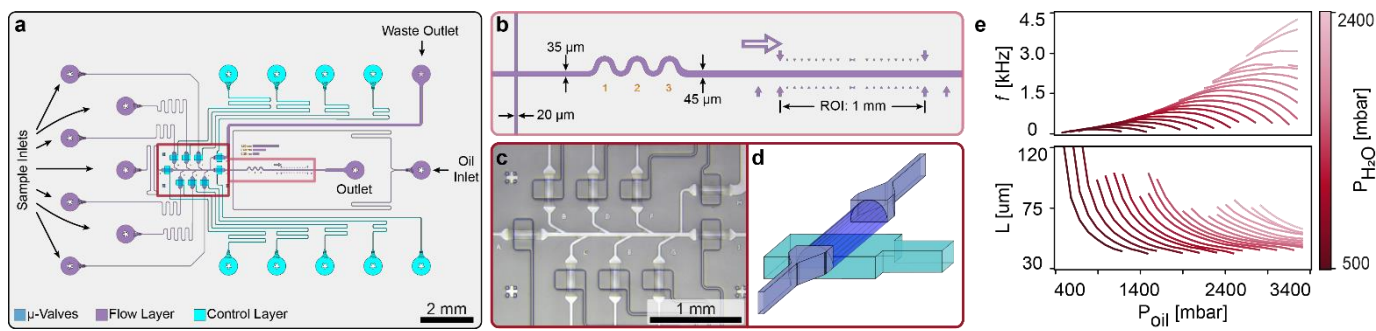

**Fig. 2. Design and characterization of droplet microfluidic chip.** (a) Design overview. (b) Zoom onto outlet channel and ROI. (c) Image of the sample inlet/valve region of an assembled droplet microfluidic chip. (d) Schematic representation of a push-up valve where the control channel is located underneath the flow channel. (e) Characterization of the droplet microfluidic chip in terms of droplet frequency and droplet length. Each curve represents a single oil phase pressure scan for different fixed aqueous phase pressures.

To test the versatility of the microfluidic platform, we characterized the droplet length and production frequency as a function of the only two tunable parameters during an experiment: the oil and aqueous phase pressures. To do so we scanned the pressure of the oil phase whilst keeping the pressure of the aqueous phase fixed; and repeated this over a range of aqueous phase pressures (Fig. 2e). The results from the parameter space sweep demonstrate that the droplet size and frequency can be quickly and easily tuned over a large range (between 45 µm to 100 µm and 100 Hz to 4.5 kHz, respectively). For most experiments reported hereafter, we targeted the production of 65 µm long droplets at a rate of 1-2 kHz. Ultimately, the chip channel geometry and dimensions determined the overall range of the droplet parameters. Nevertheless, these can be readily adjusted to fit different experimental needs.

**Real-time droplet monitoring with a multi-channel HSI microscope -** Figure 3a shows the custom-built multi-channel optical setup based around a push-broom implementation of a transmission HSI[36]-microscope, allowing us to record up to 448'000 spectra/s. The optical setup comprises three main detection modules that deliver a full characterization of a microfluidic droplet system, corresponding to an HSI, a big field of view (FoV), and a confocal-detection path. Along the HSI path, an adjustable slit in the conjugate image plane sections the illuminated FoV to a quasi-1D line, which is subsequently dispersed perpendicular to the slit by two consecutive Amici prisms. The Amici prisms play two significant roles. Firstly, they greatly reduce optical aberrations and the common distortions associated with hyperspectral imaging such as "smile" and "keystone"[37]. Secondly, they allow a straightforward way to adjust the degree of spectral dispersion by simply counter rotating the individual prisms with respect to one another [38].

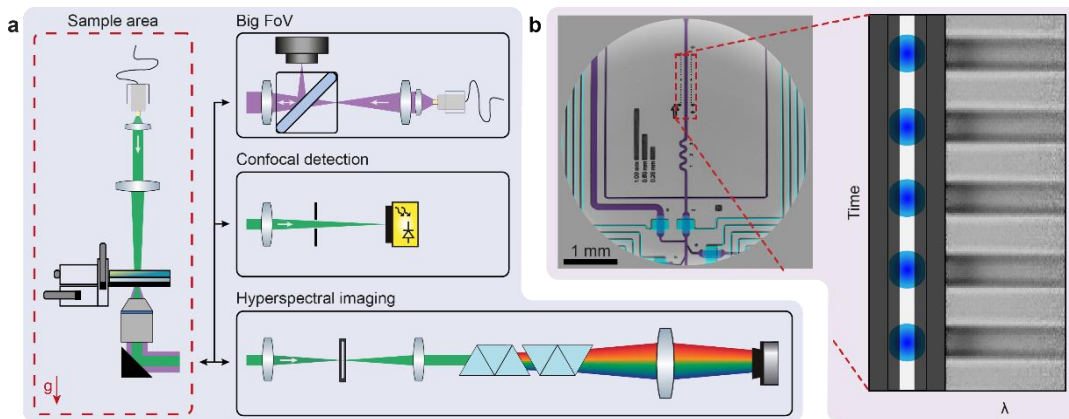

**Fig. 3. Multi-channel HSI for real-time droplet monitoring.** (a) Schematic diagram of the optical setup composed of four main modules: Sample area (g indicates the direction of gravity), big FoV, confocal detection, and HSI path. (b) Representative big FoV image of a microfluidic chip with false-color channels (light blue = control channels, blue = micro-valves, purple = flow channels). Inset: Schematic representation of the ROI as observed through a slit and corresponding hyperspectral image after dispersion of light perpendicular to the slit.

The big FoV channel is based on a reflection implementation of a partially coherent inline holography microscope. This channel aims at monitoring the operation of the microfluidic chip, ensuring the correct micro-valve actuation, and inspecting the chip for defects (Fig. 3b). Finally, for the confocal detection module a Si based avalanche photodetector (APD) records time traces of light transmitted through the droplet chip. This channel monitors the droplet production frequency at each state of an experiment.

**Self-referenced spectral read-out based on iteratively reconstructing the background** - Despite the many advantages of droplets, their curved surface leads to droplet-specific artifacts in the absorption spectra. To minimize these artifacts, the ROI featured a rectangular cross section and plug-like shaped droplets were generated. Doing so, confined the curved parts of the droplet to only the leading and trailing fronts. In other words, the central region of the droplet, corresponding to about 75% of its length, presented with a parallel surface with respect to the PDMS interface and had a uniform optical path length. To reduce scattering contributions from the droplet/oil interface, we refractive index matched (RIM) the oil phase to the aqueous phase by doping the oil phase with 1,3-Bis(trifluoromethyl)-5-bromobenzene[22,39]. Lastly to account for matrix effects, we measured and subsequently subtracted a reference spectrum[40] for each sample.

With all these factors considered, a complete dataset (i.e. sample droplets and matrix droplets) consisted of a stack of raw hyperspectral images (typically between 500 and 10'000 frames), an APD time trace, and the metadata associated with the time-dependent valve states during the measurement. The read-out is based on an iterative process that follows two main steps: droplet read-out and frequency matching. Fig. 4a depicts a schematic representation of the data analysis workflow.

The droplet readout step starts with a stack of hyperspectral images that have already been mapped to the wavelength reference system (see methods and materials; here referred to as HSI stack). Conceptually this first step iteratively reconstructs the background/reference signal from the oil segments contained within each HSI acquisition – in other words turning the HIS stack into a self-referenced absorption measurement. As an initial estimation of the background signal, the HSI stack is temporally averaged to a single background frame. Here the process of temporally averaging blurs out the droplet contribution and approximates the underlying illumination profile with respect to the oil phase. Then, based on Beer-Lambert law, an optical density (OD) data stack is calculated by dividing the initial background frame by the original HSI stack. This new OD Data stack shows increased contrast and a "flat" background. Next, the position, length, and number of droplets are detected and stored for each frame by detecting changes in OD along a line profile at a fixed wavelength range (vertical dotted line in the $n^{th}$ OD frame of Fig. 4a). Similarly, the position and extent of each oil segment is also recorded and stored for each frame.

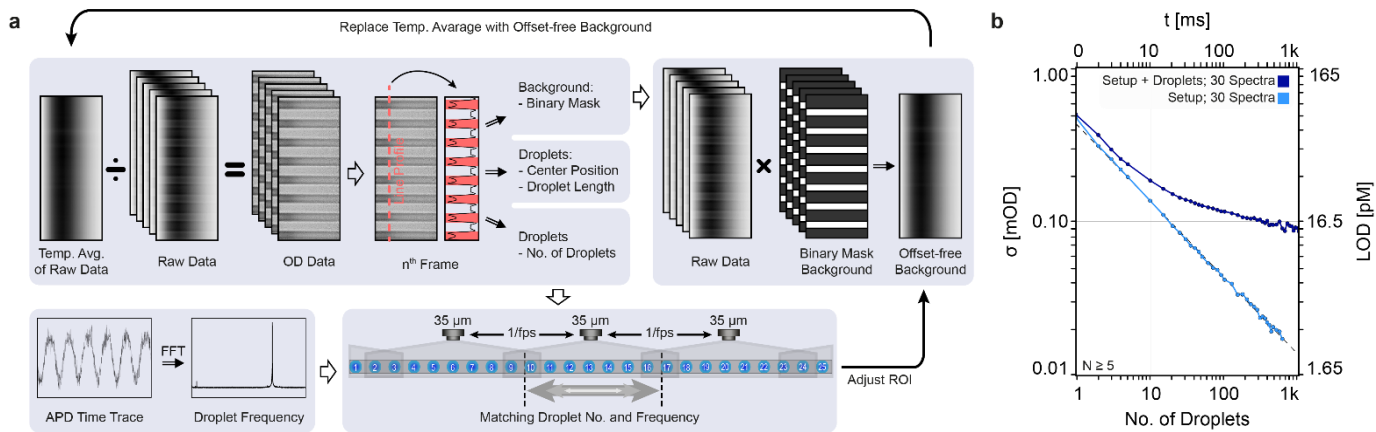

**Fig. 4. Data analysis workflow and spectral noise assessment.** (a) Schematic description of the data analysis pipeline. In short: OD data are extracted by iteratively improving the initially estimated background. The droplets themselves are localized by means of a line-profile along the spatial axis of the hyperspectral image. In combination with the APD time trace, the ROI is spatially adjusted to match the droplet production rate. (b) Average noise in the spectral region of 520-550 nm as a function of the number of averaged droplets (bottom axis) and corresponding effective acquisition time (top axis). The right y-axis represents to the corresponding LOD (3 x noise) for a 40 nm AuNPs solution.

Based on the individual oil segments, a stack of binary masks, including a significant margin towards the edges of each segment, is generated, and then multiplied frame by frame with the HSI stack. The resulting stack is subsequently condensed into a single time-averaged image which represents a new background estimate that excludes droplet contributions. Whilst the initial background has an intrinsic offset associated to the droplet contribution in each frame, this rebuilt background drastically reduces this offset artifact. A second and third iteration through this process corrects for any remaining offsets and droplet edge contributions. Finally, based on the droplet position and length, individual spectra for each droplet are read out.

Additionally, since the total number of droplets recorded in the OD data is known by now, one can match the number of analyzed droplets per frame to the droplet production frequency. The droplet production frequency is in turn extracted from the APD time trace by Fourier analysis. To match both numbers, the length of the spatial axis in the HSI stack is adjusted.

**Spectral sensitivity** - Considering all the above-mentioned points, we assessed the sensitivity of our system using droplets loaded with 40 nm spherical gold nanoparticles (AuNPs). Based on the size of the droplets, we obtained 30 individual spectra per droplet. To determine the sensitivity, we defined two noise metrics: (i) one which included all noise sources (dark shade in Fig. 4b), and (ii) another which only considered the intrinsic noise associated to the optical system (bright shade in Fig. 4b). For the latter, we calculated the difference between two successive droplets, a differential measurement, to effectively exclude all influences that are not related to the microscope itself. By doing so, the data exhibited the characteristic slope of -0.50 associated to shot-noise limited performance. The curve including all noise sources, however, deviated from the ideal shot-noise limited performance and flattened out towards higher numbers of included droplets. Temporal drifts within our system such as the light source and the physical droplet chip, are responsible for this deviation. Nevertheless, for an average of only 10 droplets our noise limit is at 0.184 mOD, dropping to 0.116 mOD for an average of 100 droplets and even to 0.086 for an average of 1000 droplets. Compared to Probst *et al.*[22] (0.13 mOD, averaged over 500 droplets recorded during 11.11 s) we achieve the same sensitivity approximately 275-fold faster, with only 40 droplets averaged, and with 20% shorter optical path lengths.

The numbers detailed above are the effective noise limit of the platform. One can now, based on these numbers, calculate the limit of detection (LOD, usually 3σ) for any system with a spectral output in the given wavelength window. Using the 40 nm gold nanoparticles as an example, a noise of 0.116 mOD (for an average over 100 droplets) corresponds to a LOD of 19.14 pM using a 20 µm channel height as the optical path length.

**Resolving fast dynamical changes with ms-resolution** - To validate the temporal resolution of the platform and show its potential to monitor fast dynamics, we simulated a fast-changing reaction kinetic system by rapidly opening and closing the microvalves within the chip. Here the microvalves determined which of the sample inlets with different AuNPs (40 nm spheres, 80 nm spheres and 25 nm x 71 nm rods) was flown into the droplet generation portion of the chip. Thus, actuating the microvalves into different valve states effectively changed the sample composition encapsulated within each droplet. To ensure a precise and fast measurement together with a ground truth of the valve state, the experiment was fully automated based on a predefined sequence of valve actuations.

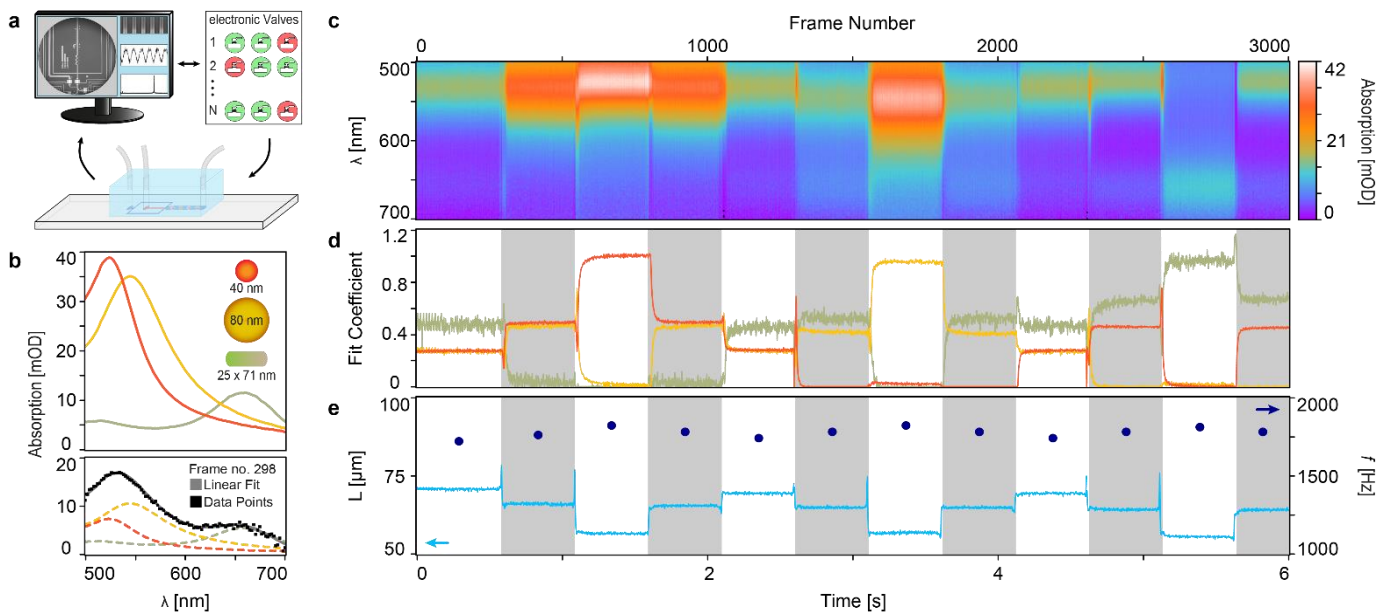

**Fig. 5. Measurement of fast spectral changes with 2 ms time resolution.** (a) Schematic representation of experiment. (b) Top: averaged absorption spectra of the three AuNPs used as a component library. Bottom: averaged droplet absorption spectra from the 298[th] frame alongst a fit to a linear combination of the described library to retrieve the individual component contributions. (c) Time-resolved absorption spectra measured for a predefined sequence of valve actuations. (d) Corresponding time-resolved composition of the droplets expressed as the linear fit coefficient retrieved from the component library derived in (b). (e) Average droplet length per frame and droplet production frequency during a specific valve state. The alternating shaded sections in (d) and (e) represent the ground truth change in the valve state.

Fig. 5b shows the highly averaged droplet spectra of individual gold nanoparticle samples which were used as a compound library. A linear combination of these three library compounds together with a flat offset to account for pressure changes due to the number of opened inlet channels was fit to each recorded spectrum. In this specific case individual droplet spectra were averaged per frame resulting in a 2 ms time-resolved absorption spectra map of the entire experiment (Fig. 5c), nevertheless if each droplet is considered individually the time resolution per spectra increases to almost 500 µs at the expense of a lower SNR. Fig. 5d shows the fit coefficients from each frame to the compound library. There, rapid changes in sample composition followed by a quick stabilization fully match with the associated changes in the valve state and the corresponding rise and fall times of the microvalves. The lower nanorod OD in the compound library led to the higher noise levels in the fit coefficient associated to this nanoparticle. Based on these fit coefficients, the effective concentration or even the absolute number of particles of each component per droplet could be determined.

Fig. 5e shows the time evolution of the average droplet length as well as the overall frequency for the system in response to changes in the valve state. Apart from a sharp peak after closing an inlet (short pulse of higher pressure due to the collapse of a flow channel) and a small dip upon opening an additional inlet (minimal pressure release due to the opening of an additional channel), the droplet size stabilized almost immediately. When the three sample inlets were open the chip produced the largest droplets. This is because the volume from several aqueous inlets competed against a constant oil pressure which allowed for a minimal longer filling phase for each droplet. Conversely, the shortest droplets were generated when only one aqueous inlet was opened. In agreement with the characterization of the droplet generation of the chip (Fig. 2e), the droplet frequency behaved exactly opposite to the droplet length. Namely, larger droplets were generated at a lower rate and vice versa.

**DNA sensing and reaction monitoring over time** - To demonstrate the versatility of our platform, we designed two experiments showcasing its capabilities as (i) a sensitive label-free biosensor, and (ii) for real-time monitoring of the evolution of complex systems. Both experiments utilize a system composed of two sets of AuNPs functionalized with different DNA capture probe strands, referred to as A-AuNP and B-AuNP, respectively (Fig. 6a). These two AuNP in the presence of a target DNA strand hybridize to form dimers and larger agglomerates in a concentration dependent manner, which in turn can be measured by changes in the absorption spectrum.

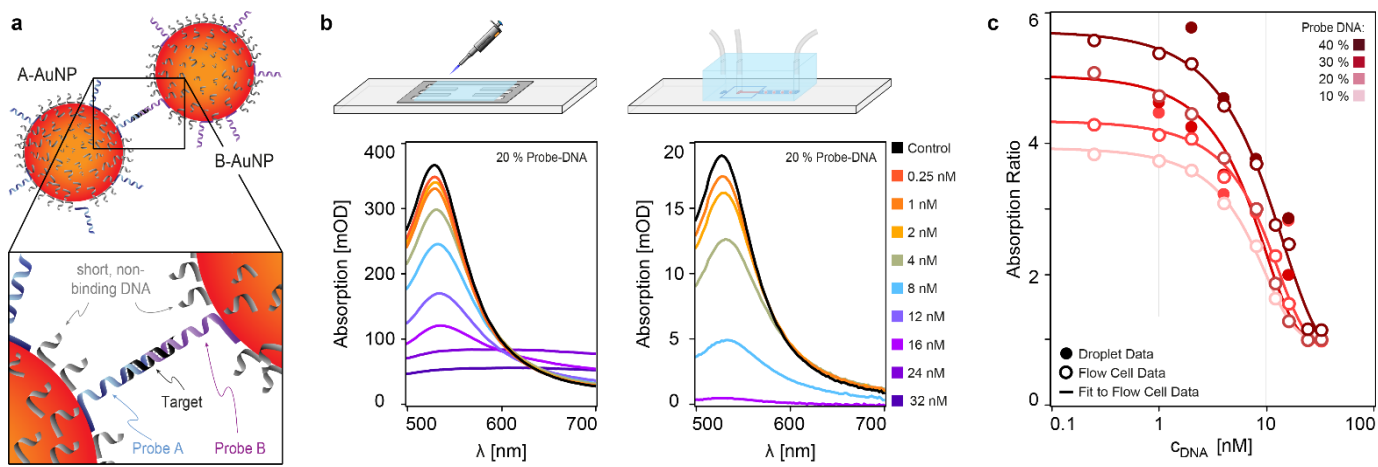

**Fig. 6. End-point sensing of DNA.** (a) Schematic representation of the DNA-AuNP sensing system. (b) End-point validation of the DNA sensor for different target DNA concentrations measured in a flow cell and in a microfluidic droplet chip. (c) Corresponding dose-response curves from (b) for different probe DNA doping percentages.

In a first experiment we validated the performance of our system as a concentration-dependent sensor by running end-point measurements in a flow cell (Fig. 6b). Here we considered an incubation time of at least 60 minutes as an end point measurement. We then repeated these measurements for selected target concentrations with our droplet microfluidic system. To account for possible non-specific interactions between the sample and the different interfaces, we quantitated the dose response as the absorption ratio between two different wavelengths (in this case $\lambda_{peak} / \lambda_{610\,nm}$). This ratiometric approach suppressed artifacts associated with considering absolute absorption values. Fig. 6c shows the corresponding dose-response curves for the data recorded in flow cells (open circles) and in droplets (filled circles) for different doping percentages of the probe DNA on the functionalized AuNPs. Thereby, the solid line represents a fit of the flow cell data points based on a sigmoidal dose response model (see material and methods). Further, the dose response curves show that by changing the capture probe doping, the linear sensing range and sensitivity of the system can be adjusted. We observed the same general trends for the droplet data. However, they did not exactly match the calibration curves obtained from the flow cells. This is expected due to the slight differences in the environment between the two systems.

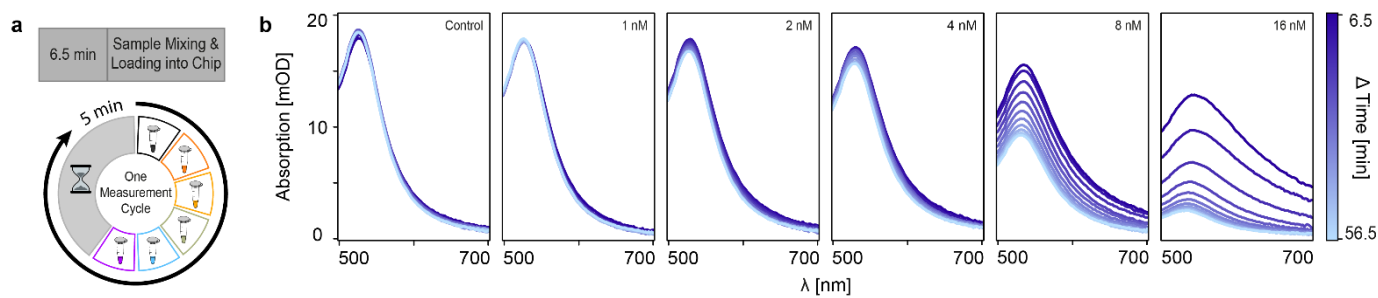

**Fig. 7. Monitoring the evolution of a dynamical system.** (a) Schematic representation of automated measurement procedure. (b) Observation of slow dynamics caused by the formation of dimers and agglomerates of DNA-AuNPs over time for different concentrations. Data acquisition started 6.5 min after mixing DNA-AuNPs and target strand. Sampling was performed every 5 min.

Using the same nanoparticle system, in the second experiment we showcase how the platform can monitor the evolution of slow dynamic processes of several samples across multiple timescales all within the same chip, spanning from minutes to an hour. For this we leverage on the platform's fully automated operation over prolonged periods in a multiplexed manner. More precisely, we monitored the response of the nanoparticle system to five different target DNA concentrations and a negative control by measuring them sequentially every five minutes for up to an hour (Fig. 7a). The first measurement cycle started 6.5 min after addition of the target strand. This period corresponds to the dead time associated with the preparation of all the samples and setting up the automated measurement. Fig. 7b shows the corresponding time evolution of each sample in response to different concentrations of target DNA. As expected, no notable change was observed for the control, whereas in the presence of the target DNA, the time-resolved spectra evolved in a target concentration dependent manner, with large changes observed at the beginning of the measurement before plateauing.

## Discussion

This work outlines a method to observe dynamical changes across different time scales in a label free manner and applicable to any solution-based sample with spectral features in the visible range. The developed platform satisfies the requirements regarding high throughput, low sample volumes and multiplexing, all whilst offering a high degree of control and versatility. In the context of fast dynamical changes, we achieve a 2-ms time resolution based on averaging all the droplets per frame, and therefore defined by the frame rate of the camera. However, when considering the spectra of each droplet individually, then our platform can provide a time resolution of less than 250 µs, which is almost a 50-fold improvement compared to similar work[30].

With respect to sensing, our approach is based on an amplification-free hybridization of a target strand with two differently functionalized AuNPs. With a target length of only 22 nucleotides, our sensor provides a route for miRNA detection[41–44] that is sensitive to single base mismatches[45]. Although the sensing approach was not optimized any further than screening different probe strand doping percentages, we demonstrated a target strand sensing range between 250 pM and 32 nM, with a roughly one order of magnitude linear range. For comparison, Pei et al.[46] used a very similar approach (13 nm AuNPs, 30 nucleotide target, bulk measurement) and observed DNA hybridization in a range between 0.5 nM and 10 nM. Hwu et al.[47] also used a dimer-based sensor (50 nm AuNPs, 24 nucleotide target) but optimized its probe strand density and measured in a darkfield configuration where the hybridized AuNPs are immobilized to the surface. By doing so, they increased the linear sensing range to approximately two orders of magnitude. Nonetheless, despite its simplicity, our sensor demonstrates great potential for diverse sensing applications and optimizing the target concentration range could further enhance its capabilities.

Our integrated optofluidic platform currently has two constraints: (i) The surface passivation within the microfluidic chips was not optimized. This is especially critical for diagnostic applications aimed at analyzing complex sample matrices such as biological fluids. However, there are surface treatment protocols to reduce nonspecific absorption[48–50] or even approaches to directly modify the PDMS towards prohibiting non-specific binding[51]. (ii) The sensitivity of our droplet measurements deviates from the ideal shot noise limit that our optical setup can deliver. Since the optical setup is shot noise limited, these deviations are connected to minute time-dependent fluctuations in the droplet and channel optical properties, which could be greatly reduced by an improved droplet referencing. This could be achieved with a droplet microfluidic system that generates alternating sample and reference droplets[52–55] – thereby suppressing all potential sources of drift over time. Unfortunately, this change requires a redesign of the microfluidic chip as well as the analysis workflow of the data.

In summary, our work showcases an optofluidic platform that combines droplet microfluidics and hyperspectral imaging with measurement automation. Regarding microfluidics, we established a simple yet very versatile and highly controllable droplet chip design. Said droplet chip provides a solid basis for potentially more complex on-chip operations and sample manipulation and eliminates the need for hydrophobic treatment of the flow channels. Regarding optical imaging, we developed a custom microscope optimized for absorption measurements. Nevertheless, due to the versatility of the system to record any spectral features in the visible range, the microscope could be easily modified for fluorescence or even Raman/SERS spectroscopic measurements. Regarding data analysis, we outlined a droplet-based workflow that allows for self-referenced absorption measurements using the oil phase as a background. In addition, we determined the minimal resolvable spectral signals and showcased how the platform can monitor dynamic systems over different timescales from ms up to hours. Finally, we presented a sensing approach for the detection of short DNA strands down to pM concentrations and highlighted the possibilities of long-term sample monitoring.

We envision our platform to unlock new opportunities, especially when combined with slowly changing dynamical systems. For instance, a promising application would be to real-time monitor the secretion profile of an organ-on-a-chip, whereby our platform could continuously study the effects of external stimuli - such as drugs, heat, and pH variations. This application could not only benefit fundamental biomedical and pharmaceutical research, but also advance personalized medicine, drug development, and even contribute to reducing animal testing.

**Materials and Methods**

**Reagents** - 3M™ Novec™ 7500 fluorinated oil containing 2 % dSURF (DR-RE-SU-30) was purchased form Fluigent, France. 10x Tris Buffered Saline (TBS, 1706435) and 10% Tween 20 Solution (1610781) were purchased from Bio-Rad, USA and mixed to TBST (1x TBS containing 0.1 % Tween 20 Solution). 1,3-Bis(trifluoromethyl)-5-bromobenzene (290157-50g), ChemiBLOCKER (2170), phosphate buffered saline (PBS, P4474-1L), Tris Acetate-EDTA buffer (TAE, F4038-1L), sodium chloride (S9888-500G), and nuclease-free water (W4502-1L) were purchased from Merck, Switzerland. Tris Acetate-EDTA buffer was diluted in autoclaved water and mixed with sodium chloride to a 1.5x TAE, 1 M NaCl buffer.

**Gold Nanoparticle** - 40 nm gold nanospheres, bare (citrate), 2.2 nM at 20 OD (AUCR40) and 80 nm gold nanospheres, bare (citrate), 230 pM at 20 OD (AUCR80) were purchased from nanoComposix, USA. Gold nanorods 25 nm x 71 nm, bare (citrate), 170 pM at 1.2 OD (A12-25-650-CIT-DIH-1-25) were purchased from Nanopartz, Canada.

**DNA strands** - All the oligos were purchased from Integrated DNA Technologies IDT, USA, and diluted to a stock concentration of 100 µM. The oligo sequences are as follows:

| Oligo sequence | 5' → 3' |
|---|---|
| Probe strand A | CTG ATA AGC TAT TTT TTT TTT TTT TTT TTT TTT TTT TTT TTA AAA AAA AAA |
| Probe strand B | AAA AAA AAA ATT TTT TTT TTT TTT TTT TTT TTT TTT TTT TTC AAC ATC AGT |
| Short DNA | AAA AAT TTT T |
| Target | TAG CTT ATC AGA CTG ATG TTG A |
| Control | ACA GAT TGA GTT AGA CTT TCG T |

**Fabrication of molds for microfluidic chips** - Individual molds for flow and control layer were fabricated by means of soft lithography using direct laser writing (µMLA, Heidelberg Instruments, Germany) on a 10-in silicon wafer (WSM4052525XB1314SNN1, MicroChemicals GmbH, Germany). The mold of the control layer was fabricated through a single exposure of 20 µm SU8 1060 negative photo resist (Gersteltec Engineering Solutions, Switzerland). The flow layer mold, on the other hand, required a two-step fabrication process to simultaneously allow for semi-rounded (valve areas) and squared (all other areas) channel cross sections. To achieve this, the wafer was first structured with SU8 1060 photo resist as described above. Then, the same wafer was coated again with 20 µm of AZ P4620 positive photo resist (MicroChemicals GmbH). Before structuring the photo resist by a second exposure, the design was aligned with the pre-existing SU8. A final baking step causes the AZ structures to slightly melt and reshape into semi-rounded cross sections whilst the SU8 photo resist underwent a glass transition but maintains the squared channel profile. Finally, both molds were silanized with chlorothrimethylsilane (92360, Sigma Aldrich, Switzerland) at low pressure to improve longevity.

**Fabrication of microfluidic PDMS chip** - To produce the two-layer microfluidic droplet chips, the structures from the two molds were transferred to PDMS (Sylgard 184, Dow Chemical, USA) mixed at 10:1 w/w polymer to curing agent ratio. For the flow layer, the PDMS was drop-cast onto the respective mold to a height of approximately 5 mm. To make the control layer, a small amount of PDMS was spin-coated onto the mold at 2000 rpm resulting in a layer thickness of about 40 µm. Both molds were then placed under vacuum to degas until bubbles were no longer visible (typically around 90 min). Afterwards, the molds were baked in a convection oven at 80 °C for one hour. Once cured, the PDMS on the flow layer mold was peeled off the substrate, individual chips were cut out and inlet holes were punched. Then, the individual flow layer chips and the control layer mold (still covered with 40 µm of PDMS) were exposed to oxygen plasma (10s, 300 W, 8 sccm, Atto Low Pressure Plasm System, Diener Electronics, Germany) before manual alignment under a stereo macroscope (Z16 APO, Leica, Germany). For a stronger bond between the two layers, the aligned chips were baked again for 15 min at 80 °C in the oven. The chips were then cut off from the control mold and the inlet holes for the control layer were punched. To finish the assembly, the two-layer PDMS structure and microscope glass slides (17204894, Erpedia, USA) were exposed to oxygen plasma of the same conditions as before, bound together, and baked for another 15 min at 80 °C in the oven.

**Flow cell fabrication** - Flow cells were used as simple and fast alternative to droplet microfluidic chips to test the performance of the optical platform and to exclude droplet-specific contributions. Each flow cell consisted of twelve independent straight channels (width = 1.4 mm) open to atmospheric pressure on both ends. This design allowed for capillary-force driven loading of volumes down to 2 µL into each channel without any external equipment. The flow cells were fabricated from a microscope glass slide (17204894, Erpedia) and a cover glass (0101030, Paul Marienfeld GmbH & Co. KG, Germany) separated by a silicon spacer (500 µm, CWS-S-0.5, Grace Bio-Labs, USA). The assembly was solely based on surface energy. The channels were cut out of the silicon spacer by a cutting plotter (Cameo 4, Silhouette, USA).

**Interfacing microfluidic chips with lab equipment** - The sample inlets of the microfluidic chips were connected to Tygon® tubing ($\o_{in}$ = 0.508 mm, $\o_{out}$ = 1.524 mm, Saint-Gobain, France) via a bent metal pin (Tube AISI 304, 0.65 / 0.35 x 17.5 mm, Unimed SA, Switzerland). The tubing from the flow layer was connected to two types of sample reservoirs based on their size: i) for large volumes, 1.5 mL Eppendorf tubes interfaced with a P-Cap (P-CAP2-HP, Fluigent) or, ii) for volumes in the tens of µL, dispensing tips (LL ½" ID 0.34 mm, GONANO Dosiertechnik GmbH, Austria) with a Male Luer Lock to Barb Adapter (Darwin Microfluidics, France). Both types of reservoirs were connected to pressure controllers (LU-FEZ-2000 or LU-FEZ-7000, Fluigent) with 4 mm x 2.5 mm polyurethane tubing (917-2407, RS Pro, GB). The pressure controllers were operated both off-line, and programmatically via a computer by using a LineUP Link module (LU-LNK-002, Fluigent). The inlets to the control layer were connected to the same metal pins and liquid filled Tygon® tubing as described above. The tubing from the control layer was then interfaced to a custom-built electronic valve unit that was programmatically controlled by a computer.

**Passivation of microfluidic tubing** - Passivation of the microfluidic Tygon® tubing - if performed - was achieved by first rinsing the tubing with a 10-fold dilution of ChemiBLOCKER in TBST. Then, the tubing was incubated with the same reagent for a few minutes before being thoroughly rinsed with PBS. Finally, the tubing was flushed with nitrogen for a few seconds.

**Microscope** - The custom-built microscope follows a flexible and modular approach combining three optical paths:

(i) The HSI path is based on a transmission microscope. For illumination a 554 nm fiber coupled light emitting diode (MINTF4, Thorlabs, USA) was connected to a 550 µm multimode fiber (M37L02, Thorlabs). The output of the fiber was imaged onto the sample by a telescope comprising a $f = 11$ mm aspheric lens (C397TMD-A, Thorlabs) and a $f = 40$ mm achromatic doublet lens (AC254-040-A, Thorlabs) arranged in a 4f configuration. The sample was positioned onto a custom built sample holder built on top of xyz-piezo translation stage (nanoCube, Physik Instrumente GmbH, Germany) mounted on top of a manual xyz-translations stage with a range of 25 mm x 25 mm x 12.5 mm. Light from the sample was collected by an 4x, 0.2 NA objective (TL4S-SAP, Thorlabs) before passing through a dichroic mirror (long pass 490 nm, DMLP490L, Thorlabs) and imaged onto a slit (width = 50 µm, VA100/M, Thorlabs) to produce a quasi 1D image. A second $f=150$ mm achromatic doublet lens (AC508-150A, Thorlabs) relay imaged the BFP of the objective onto two subsequent but counter-rotated Amici prisms (117240, Equascience, France) to spectrally disperse the light perpendicular to orientation of the slit. To allow for a multichannel readout a 90:10 (R:T) beam splitter (BSX16, Thorlabs) was introduced before the Amici prisms. Finally, the light reflected from the beamsplitter and dispersed by the prisms was imaged onto a CMOS camera (GS3-U3-23S6H, Sony IMY174 Sensor, 1200 x 1920 pixel, 5.86 µm x 5.86 µm pixel size, Point Grey, FLIR Systems, USA) by a $f = 200$ mm achromatic doublet lens (AC508-200-A, Thorlabs). This channel provided an effective 4x magnification (i.e. 1.488 µm per pixel) and an average spectral dispersion of 0.782 nm / pixel between 504 - 613 nm.

(ii) The confocal detection path corresponds to the portion of light transmitted through the above-described 90:10 beamsplitter. This light was focused using a $f = 150$ mm achromatic lens (LA1433-A, Thorlabs) onto an APD (APD120A2/M, Thorlabs) with a 50 µm pinhole (P50K, Thorlabs) mounted right in front of it. All optical elements were arranged in a 4f configuration with respect to the slit, leading to an effective 3x magnification.

(iii) The big FoV path corresponds to a spatially incoherent digital holographic optical system which was based on a common-path microscope operating in reflection with all optical elements arranged in a 4f configuration. For the illumination, a 415 nm fiber coupled light emitting diode (M415F3, Thorlabs) was connected to a 600 µm multimode fiber (M29L02, Thorlabs). The light outcoupled from the fiber by a $f = 6.2$ mm aspheric lens (C171TMD-A, Thorlabs) passed through a 1:1 relay-imaging system formed by two $f = 125$ mm achromatic doublet lenses (AC254-125-A, Thorlabs). In between the relay imaging system, a 50:50 beamsplitter (BSW27, Thorlabs) was placed to separate the illumination and imaging paths. This module was coupled to the rest of the optical microscope via a dichroic mirror (long pass 490 nm, DMLP490L, Thorlabs). Light reflected off from the dichroic mirror was focused onto the sample using the same 4x, 0.2 NA objective (TL4S-SAP, Thorlabs) as before. In the imaging arm, light from the sample was collected by the same objective and was subsequently imaged onto a CMOS camera (a2A4504-18umPRO – Basler ace 2, 4504 x 4504 pixel, 2.74 µm x 2.74 µm pixel size, Basler AG, Germany) upon reflecting off the 50:50 beamsplitter.

**Calibration of hyperspectral camera (mapping pixels to wavelength)** - The HSI path of the system provides images with both spatial (y-axis) and spectral (x-axis) information. However, the information is reported according to the reference system of the camera: pixels. To read-out spectral information, a coordinate transformation from pixels to wavelength is required. We achieved this by a calibration step which used a total of nine filters (FF02-472/30-25, Semrock, USA, and MF497-16, FL532-10, FB570-10, FLH635-10, FB650-10, FB700-10, FL740-10, FGB67, Thorlabs) with known and sharp absorption peaks. Said filters were mounted in front of the light source (before the sample stage) and imaged along the HSI path and compared against a reference spectrometer (USB4000, Ocean Optics, Ocean Insight, USA). The recorded data was then temporally averaged (2000 frames recorded over 40 s for the hyperspectral images, and 100 scans for the reference spectrometer) to increase the signal to noise ratio. For each filter, the wavelength of the absorption peak was found in the spectrometer data. In the HIS channel, the pixel position of the absorption peak along the spectral axis (x-axis) was located along each spatial coordinate (y-axis). Finally, the detected pixel-positions were mapped to the corresponding wavelengths and fit to a $3^{rd}$ order polynomial. The resulting fit coefficients were then used, to transform the coordinates of the raw hyperspectral images. Note that by doing this, we automatically accounted for smile-aberrations.

**Spectral window and data binning** - In the current implementation, the platform reads spectra between 490-700 nm. The lower and upper limit were set by the long pass 490 nm dichroic mirror in the optical setup, and the spectra of the light source itself, respectively. Previous tests have shown that for a shot noise limited and fixed photon budget system, the SNR for reading out spectral features does not improve when sampling beyond the Nyquist criteria. Therefore, we chose a spectral resolution of 2 nm by pixel binning along the spectral axis of the transformed HSI data.

**Automated Measurements based on .txt file sequences** - Our microfluidic chips were equipped with micro-valves able to close flow channels pressurized up to 80 % of the applied pressure to the control layer. Thus, a fully pressurized microfluidic chip could be operated solely based on actuating the required microvalves which means that the entire microfluidic operation of the system could be programmatically controlled. To do so, we implemented a producer/consumer queue structure in LabView (2020, National Instruments, USA) to perform a set of specific tasks based on the four commands "open channel x", "close channel x", "wait" and "measure". These tasks were then pre-programed into a .txt- file which then could be uploaded and sent to the electronic valves and camera respectively. A typical measurement of a single sample would thereby look as follows and usually last between 20 and 30 seconds: "open channel x" > "open channel waste" > "wait" > "close channel waste" > "open channel main" > "wait" > "measure" > "close channel main" > "close channel x" > "open channel matrix" > "open channel waste" > "wait" > "close channel waste" > "open channel main" > "wait" > "measure" > "close channel main" > "close channel matrix".

**Droplet chip characterization** - For the chip characterization droplets composed of milli-Q water were produced under different pressure conditions of the oil and aqueous phase, respectively. Specifically, the oil phase pressure was scanned over the entire range that allowed droplet production (starting at 3300 mbar, going down to 500 mbar, step size of 100 mbar) whilst keeping the aqueous phase pressure fixed. By iteratively scanning, an aqueous phase pressure range between 2400 mbar down to 500 mbar in steps of 100 mbar was covered. For each condition 1000 frames were recorded over a time of 2 s. The droplet frequency and size were then determined from all the droplets recorded during the acquisition.

**Noise Assessment** - For the noise assessment, a total of 55'775 droplets containing a solution of 40 nm AuNPs (production frequency = 1030 Hz, average length = 86.5 ± 1.7 µm, frequency not matched to ROI) were recorded. From this data set, all droplets detected in the center region of the ROI (length = 104.1 µm, 6'056 AuNP droplets, 30 spectra per droplet) were further processed. After averaging a certain number of spectra (i.e. droplets), the average and standard deviation (STD) of the absorption were determined for each wavelength. The average value of the STD for the spectral region between 520-550 nm (encloses most relevant peaks) was then used as a metric for the noise. Spectra were averaged in two ways. (i) Averaging droplet spectra. This approach includes all noise sources present in the system – including drifts, light fluctuations, chip aging, optical noises, etc. (ii) Averaging the difference between two consecutive droplet spectra. This differential approach excludes most noise sources and tests whether the optical system has shot noise limited performance.

**Experimental procedure and data analysis for fast dynamical changes** - For this experiment, 40 nm and 80 nm spherical AuNPs stock solutions were used. The nanorods, however, were concentrated approximately 5-fold by centrifugation. The experiment was fully automated in regard to the sequence of valve activation. Each valve state was active for only 500 ms, except for the first and last one due to manual camera triggering. Droplets were recorded with a frame rate of 500 Hz and 20 individual spectra were considered per droplet to increase SNR. For Fig. 5c – e, all droplets within a frame were averaged to produce a single spectrum per frame. For the droplet library presented in Fig. 5b, each sample was recorded independently during 2 s and all recorded droplets were averaged to a single spectrum.

**DNA-AuNP preparation** - AuNPs were functionalized with DNA according to the microwave assisted heating-drying protocol[56]. In short, equal volumes of 40 nm AuNP at 2.2 nM and DNA at 10 µM were mixed in a small glass vial. The DNA component comprised a mixture of short DNA doped with a percentage of either probe strand A or probe strand B. The complete volume of this mixture was typically kept below 120 µL to facilitate evaporation. The glass vial was then placed in a microwave oven at a power of 1000 W until the liquid was completely evaporated – typically around 20 min. Afterwards, the remaining AuNPs in the glass vial were resuspended in 300 µL nuclease free water and transferred to a 500 µL DNA LoBind Eppendorf tube (0030108035, Eppendorf, Switzerland). The DNA-AuNPs were then washed three times by centrifugation (4700 rpm or 1400 g, mySPIN™ 12 mini centrifuge, Thermo Fisher Scientific, Switzerland).

**DNA sensing** - Before using any of the DNA functionalized AuNPs, the concentration of each particle and doping type was equilibrated to a common starting value. Then both probe strand functionalized AuNPs were mixed with either target or control DNA (in 1.5x TAE with 1 M NaCl) in a 1:1:2 v/v/v ratio and incubated for at least 60 min at room temperature. Acquisition of all droplet data was fully automated whereby each sample was sequentially measured (1000 frames over 2 seconds; 35 µs exposure time) from highest to lowest concentration. The production frequency reached up to 1.5 kHz (for 30 % and 40 % doping) and 2.1 kHz (for 20 % doping). All recorded droplets were averaged to generate a single spectrum. Flow cell data were recorded over 5 s with a frame rate of 100 Hz. During the acquisition, the stage was manually moved to avoid imprinting static defects. All recorded HSI frames were temporally averaged and referenced to a background measurement. At nine equally spaced positions along the spatial axis, 30 spectra were read-out and averaged to a single spectrum. To quantify the degree of agglomeration, the absorption ratio $\lambda_{peak} / \lambda_{610\,nm}$ was chosen as a metric for all data sets. For the flow cell data, the data point observed with the above-mentioned metric were fit to the following sigmoidal dose response model: $f(x) = a_{sat} + \{(a_0 - a_{sat}) / (1 + 10^{([center] - [DNA])p})\}$, where $a_{sat}$ and $a_0$ are the top and bottom asymptotes, *[center]* is the target concentration at the center of the curve, *[DNA]* is the concentration of the target and *p* is the hill slope. The four fits presented with $R^2$ values of 0.994, 0.996, 0.995 and 0.998 for 40 %, 30 %, 20% and 10 % of probe strand doping.

**Reaction Monitoring** - For reaction monitoring, the samples described under "DNA sensing" were not allowed to incubate but loaded into the droplet chip immediately after mixing. Acquisition of all droplet data was fully automated. Specifically, each sample was sequentially measured (1000 frames over 2 seconds; 35 µs exposure time) from highest to lowest concentration every five minutes. Droplet length was below 70 µm, whilst the production frequency reached up to 1.5 kHz. All recorded droplets for a given time point were averaged to generate a single spectrum.

## Acknowledgments

The authors would like to thank Jose Garcia Guirado for helpful discussions regarding microfluidic chip design and fabrication.

**Funding:**
No external funding.

**Author contributions:**
Conceptualization: JOA, RQ
Methodology: MS, JOA
Investigation: MS, JOA
Software: MS, JOA

Formal analysis: MS
Visualization: MS
Supervision: JOA, RQ
Writing—original draft: MS
Writing—review & editing: MS, JOA, RQ

**Competing interests:**
Authors declare that they have no competing interests.

**Data and materials availability:**
All data needed to evaluate the conclusions in the paper are present in the paper and/or the Supplementary Materials.

**Supplementary Materials**
none